\documentclass[a4paper, 12pt, onecolumn]{article}
\pdfoutput=1

\usepackage[cmex10]{amsmath}	
\usepackage{amssymb}
\usepackage{amsthm}

\usepackage{mathtools}	
\mathtoolsset{showonlyrefs}
\usepackage{thmtools}

\usepackage[T1]{fontenc}
\usepackage[latin1]{inputenc}
\usepackage{lmodern}

\usepackage[english]{babel}

\newcommand{\RR}{\ensuremath{\mathbb{R}}}
\renewcommand{\P}{\ensuremath{\mathcal{P}}}	
\renewcommand{\S}{\ensuremath{\mathcal{S}}}	

\newcommand{\X}{\ensuremath{\mathcal{X}}}
\newcommand{\W}{\ensuremath{\mathcal{W}}}

\renewcommand{\vec}[1]{\ensuremath{\boldsymbol{#1}}}	
\newcommand{\mat}[1]{\ensuremath{\MakeUppercase{\boldsymbol{#1}}}}	
\newcommand{\TODO}[1]{\textbf{TODO}}

\usepackage[nolist]{acronym}

\begin{acronym}
	\acro{PPM}{Prediction by Partial Matching}
	\acro{DMC}{Dynamic Markov Coding}
	\acro{CTW}{Context Tree Weighting}
	\acro{PAQ}{``Pack''}
	\acro{AC}{Arithmetic Coding}
	\acro{CM}{Context Mixing}
	\acro{CG}{Conjugate Gradient}
	\acro{KT}{Krichevsky-Trofimov}
	\acro{iid}{independent identically distributed}
	\acro{BWT}{Burrows-Wheeler-Transform}
	\acro{BFGS}{Broyden-Fletcher-Goldfab-Shanno}
	\acro{KKT}{Karush-Kuhn-Tucker}
	\acro{WFC}{Weighted Frequency Counting}
	\acro{MTF}{Move-to-Front}
	\acro{LP}{Laplace}
	\acro{SAKDC}{Swiss Army Knife Data Compression}
	\acro{SQP}{Sequential Quadratic Programming}
	\acro{bpc}{bits per character}
  \acro{OGD}{Online Gradient Descent}
  \acro{BW}{Beta-Weighting}
  \acro{pd}{probability distribution}
\end{acronym}

\def\Compact{}
\usepackage[margins=normal,bibliography=normal]{savetrees} 

\ifdefined\Compact{}    
    
  \renewenvironment{thebibliography}[1]{%
    \begin{oldthebibliography}{#1}%
      \small
      \setlength{\parsep}{0pt}%
      \setlength{\itemsep}{-1pt}%
  }{%
     \end{oldthebibliography}%
  }

  \newenvironment{prf}[1][Proof]
  {\vspace{-0.75\topsep}\begin{proof}[#1]}{\end{proof}\vspace{-.75\topsep}}
  
  \newtheoremstyle{plain1}
    {0.3\topsep}   
    {0.3\topsep}   
    {\itshape}  
    {}       
    {\bfseries} 
    {.}         
    {5pt plus 1pt minus 1pt} 
    {}          
  \newtheoremstyle{definition1}
    {0.3\topsep}   
    {0.3\topsep}   
    {\normalfont}  
    {}       
    {\bfseries} 
    {.}         
    {5pt plus 1pt minus 1pt} 
    {}          
  \newtheoremstyle{remark1}
    {0.3\topsep}   
    {0.3\topsep}   
    {\normalfont}  
    {}       
    {\itshape} 
    {.}         
    {5pt plus 1pt minus 1pt} 
    {}          

  \newcommand{\plaintheoremstyle}{\theoremstyle{plain1}}
  \newcommand{\remarkstyle}{\theoremstyle{remark1}}
  \newcommand{\definitionstyle}{\theoremstyle{definition1}}
    
  \AtBeginDocument{
    \setlength{\abovedisplayskip}{6pt}
    \setlength{\abovedisplayshortskip}{6pt}
    \setlength{\belowdisplayskip}{6pt}
    \setlength{\belowdisplayshortskip}{6pt}
    
    \setlength{\floatsep}{6pt}
    \setlength{\textfloatsep}{6pt}
    \setlength{\intextsep}{6pt}
  }
\else
  \newenvironment{prf}[1][Proof]{\begin{proof}[#1]}{\end{proof}}
  \newcommand{\plaintheoremstyle}{\theoremstyle{plain}}
  \newcommand{\remarkstyle}{\theoremstyle{remark}}
  \newcommand{\definitionstyle}{\theoremstyle{definition}}
\fi

\usepackage{booktabs}
\usepackage[top=1in, left=1.25in, textwidth=6in, textheight=9in]{geometry} 	
\usepackage[ruled,linesnumbered,vlined]{algorithm2e}

\plaintheoremstyle
\newtheorem{theorem}{Theorem}[section]
\newtheorem{proposition}[theorem]{Proposition}
\newtheorem{lemma}[theorem]{Lemma}

\remarkstyle
\newtheorem{remark}[theorem]{Remark}

\definitionstyle
\newtheorem{definition}[theorem]{Definition}
\newtheorem{example}[theorem]{Example}
\newtheorem{observation}[theorem]{Observation}

\newcommand{\T}{\ensuremath{^\mathsf{T}}}

\newcommand{\eps}{\ensuremath{\varepsilon}}

\newcommand{\powlw}[2][w]{\ensuremath{{2^{-\vec #1{}\T\vec l(#2)}}}}
\newcommand{\pmin}{\ensuremath{p_{\mathrm{min}}}}
\newcommand{\pmax}{\ensuremath{p_{\mathrm{max}}}}

\newcommand{\mix}{\textsc{mix}}
\newcommand{\best}{\textsc{best}}
\newcommand{\mixogd}{\textsc{mix-ogd}}
\newcommand{\geo}{\textsc{geo}}

\newcommand{\lin}{\textsc{lin}}
\newcommand{\linogd}{\textsc{lin-ogd}}

\newcommand{\proj}{\ensuremath{{\mathrm{proj}}}}

\title{\Large{Linear and Geometric Mixtures -- Analysis}}
\author{
	\vspace{2mm}
	Christopher Mattern \\ 
	Technische Universität Ilmenau\\
	Ilmenau, Germany \\
	\texttt{christopher.mattern@tu-ilmenau.de}
}
\date{}	

\begin{document}

\par\noindent\textbf{This paper is a preprint (IEEE ``accepted'' status).}

\bigskip\par\noindent
\textbf{IEEE copyright notice.}~\mbox{\copyright{}} \the\year{} IEEE. Personal use of this material is permitted. Permission from IEEE must be obtained for all other uses, in any current or future media, including reprinting/republishing this material for advertising or promotional purposes, creating new collective works, for resale or redistribution to servers or lists, or reuse of any copyrighted component of this work in other works.	
\newpage

\maketitle
\thispagestyle{empty}	



\begin{abstract}
\noindent
Linear and geometric mixtures are two methods to combine arbitrary models in data compression. Geometric mixtures generalize the empirically well-performing \acs{PAQ}7 mixture. Both mixture schemes rely on weight vectors, which heavily determine their performance. Typically weight vectors are identified via \acl{OGD}. In this work we show that one can obtain strong code length bounds for such a weight estimation scheme. These bounds hold for arbitrary input sequences. For this purpose we introduce the class of \emph{nice} mixtures and analyze how \acl{OGD} with a fixed step size combined with a nice mixture performs. These results translate to linear and geometric mixtures, which are nice, as we show. The results hold for \acs{PAQ}7 mixtures as well, thus we provide the first theoretical analysis of \acs{PAQ}7.
\end{abstract}


\section{Introduction}\label{sec:introduction}

\paragraph{Background.}
The combination of multiple probability distributions plays a key role in modern statistical data compression algorithms, such as \ac{PPM}, \ac{CTW} and \ac{PAQ} \cite{cm_dcc2012,hbdc,ppm_ii,ctw_95}. Statistical compression algorithms split compression into \emph{modeling} and \emph{coding} and process an input sequence symbol-by-symbol. During modeling a model computes a \emph{model distribution} $p$ and during coding an encoder maps the next character $x$, given $p$, to a codeword of a length close to $-\log p(x)$. Decoding is the very reverse: Given $p$ and the codeword the decoder restores $x$. \ac{AC} is the de facto standard en-/decoder, it closely approximates the ideal code length \cite{eoit}. All of the aforementioned algorithms combine (or \emph{mix}) multiple model distributions into a single model distribution in each step. \ac{PAQ} is able to mix \emph{arbitrary} distributions. As its superior empirical performance  shows, mixing arbitrary models is a promising approach.

\paragraph{Previous Work.}
To our knowledge there exist few compression algorithms which combine arbitrary models. Volf's Snake- and Switching-Algorithms \cite{thesis_volf} were the first approaches to combine just \emph{two} arbitrary models. Kufleitner et al. \cite{cm_cmidc} proposed \acl{BW}, a \ac{CTW}-spin-off, which mixes arbitrary models by weighting the model distributions linearly. The weights are posterior probabilities on the models (based on a given prior distribution).
Another linear weighting scheme was introduced by Veness \cite{veness12}, who transferred techniques for tracking from the online learning literature to statistical data compression. His weighting scheme is based on a cleverly chosen prior distribution, which enjoys good theoretical guarantees.
Starting in 2002 Mahoney introduced \ac{PAQ} and its successors \cite{hbdc}, which attracted great attention among practitioners. \ac{PAQ}7 and its follow-ups combine models for a binary alphabet via a nonlinear ad-hoc neural network and adjust the network weights by \ac{OGD} with a fixed step size \cite{hbdc}. Up to 2012 there was no theoretical justification for \ac{PAQ}7-mixing. In \cite{cm_dcc2012} we proposed geometric (a non-linear mixing scheme) and linear mixtures as solutions to two weighted divergence minimization problems. Geometric mixtures add a sound theoretical base to \ac{PAQ}7-mixing and generalize it to non-binary alphabets. Both mixture schemes require weights, which we estimate via \ac{OGD} with a fixed step size. 

In machine learning online parameter estimation via \ac{OGD} and its analysis is well understood \cite{plg} and has a variety of applications, which closely resemble mixture-based compression. Hence we can adopt machine learning analysis techniques for \ac{OGD} in data compression to obtain theoretical guarantees. This work draws great inspiration from Zinkevich \cite{zinkevich}, who introduced projection-based \ac{OGD} in online learning and from Bianchi \cite{bianchi99} and Warmuth \cite{nnloss} who analyzed \ac{OGD} (without projection) in various online regression settings.

\paragraph{Our Contribution.}
In this work we establish upper bounds on the code length for linear and geometric mixtures coupled with \ac{OGD} using a fixed step size for weight estimation. The bounds show that the number of bits wasted w.r.t. a desirable competing scheme (such as a sequence of optimal weight vectors) is small. These results directly apply to \ac{PAQ}7-mixing, since it is a geometric mixture for a binary alphabet and typically uses \ac{OGD} with a fixed step size for weight estimation. Thus we provide the first theoretical guarantees for \ac{PAQ}. To do so, in Section \ref{sec:nice_mixtures} we introduce the class of nice mixtures which we combine with \ac{OGD} with a fixed step size and establish code length bounds. It turns out that the choice of the step size is of great importance. Next, in Section \ref{sec:geo_lin_bounds} we show that linear and geometric mixtures are nice mixtures and apply the results of Section \ref{sec:nice_mixtures}. Finally in Section \ref{sec:conclusion} we summarize our results.

\section{Preliminaries}

\paragraph{Notation.}
In general, calligraphic letters denote sets, lowercase boldface letters indicate column vectors and boldface uppercase letters name matrices. 
The expression $(a_i)_{1\leq i\leq m}$ expands to $(a_1~a_2~\dots~a_m)\T$ where ``$\T$'' is the transpose operator; the $i$-th component of a vector $\vec a$ is labeld $a_i$ and its squared euclidean norm is $\lvert\vec a\rvert^2 = \vec a\T\vec a$. By $\vec e_i$ we denote the $i$-th unit vector and $\vec 1$ is $(1~1~\dots~1)\T\in\RR^m$.
For any bounded set $\W\subset\RR^m$ let $\lvert\W\rvert := \sup_{\vec a, \vec b\in\W} \lvert \vec a-\vec b\rvert$.
Further, let $\S := \{\vec a\in\RR^m\mid \vec a\geq0\text{ and }\vec1\T\vec a=1\}$ (unit $m$-simplex).
Let $\X:=\{1, 2,\dots, N\}$ be an alphabet of cardinality $1<N<\infty$ and let $x_a^b := x_a x_{a+1} \dots x_b$ be a sequence over $\X$ where $x^n$ abbreviates $x_1^n$.
The set of all probability distributions over $\X$ with non-zero probabilities on all letters is $\P_+$ and with probability at least $\eps>0$ on all letters is $\P_\eps$.
For $p_1, p_2, \dots, p_m\in\P\subseteq\P_+$ let $\vec p(x) = (p_i(x))_{1\leq i\leq m}$ be the vector of probabilities of $x$, the matrix $\mat P := (\vec p(1)~\dots~\vec p(N))$ is called a probability matrix over $\P$.
Furthermore we set $\pmax(x; \mat P) := \max_{1\leq i\leq m} p_i(x)$ and $\pmax(\mat P) := \max_{x\in\X} \pmax(x; \mat P)$; $\pmin(x; \mat P)$ and $\pmin(\mat P)$ are defined analogously. We omit the dependence on $\mat P$, whenever clear from the context.
The natural logarithm is ``$\ln$'', whereas ``$\log$'' is the base-two logarithm. For a vector $\vec a$ with positive entries we define $\log \vec a := (\log a_i)_{1\leq i\leq m}$.
For $x\in\X$ and $p\in\P_+$ we denote the (ideal) code length of $x$ w.r.t. $p$ as $\ell(x, p) := -\log p(x)$.
The expression $\nabla_{\vec w} f := (\partial f/\partial w_i)_{1\leq i\leq m}$ denotes the gradient of a function $f$, when unambigous we write $\nabla f$ in place of $\nabla_{\vec w} f$.

\paragraph{The Setting.}
Recall the process of statistical data compression for a sequence $x^n$ over $\X$ (see Section \ref{sec:introduction}), which we now formally refine to our setting of interest. Fix an arbitrary step $1\leq k\leq n$. First, we represent the $m>1$ model distributions $p_1, \dots, p_m\in\P_+$ (which may depend on $x^{k-1}$ and typically vary from step to step) in a probability matrix $\mat P_k$. One can think of $x^n$ and the sequence $\mat P^n := \mat P_1, \dots, \mat P_n$ of probability matrices over $\P_+$ as fixed. On the basis of $\mat P_k$ we determine a mixture distribution (for short \emph{mixture}) $\mix(\vec w,\mat P_k)$ for coding the $k$-th character $x_k$ in $\ell(x_k,\mix(\vec w, \mat P_k))$ bits. The mixture depends on a parameter vector or \emph{weight vector} $\vec w=\vec w_k$ which is typically constrained to a domain $\W$ (a non-empty, compact, convex subset of $\RR^m$). Based on an initial weight vector $\vec w_1$ (chosen by the user) we generate a sequence of weight vectors $\vec w_2, \vec w_3, \dots$ via \ac{OGD}: In step $k$ we adjust $\vec w_k$ by a step towards $\vec d := -\alpha \nabla_{\vec w} \ell(x_k, \mix(\vec w, \mat P_k))$  where $\alpha>0$ is the step size. The resulting vector $\vec v = \vec w_k+\vec d$ might not lie in $\W$, the operation $\proj(\vec v; \W) := \arg\min_{\vec w\in\W} \lvert \vec v-\vec w\rvert^2$ maps a vector $\vec v\in\RR^m$ back to the feasible set $\W$ and we obtain $\vec w_{k+1}=\proj(\vec v; \W)$.
Algorithm \ref{alg:mixgd} summarizes this process. Next we define the general term mixture as well as linear and geometric mixtures.
\begin{algorithm}[t]  
  \SetKwInOut{Input}{Input}
  \SetKwInOut{Output}{Output}
  \SetKwInOut{Dummy}{Out\hspace{1pt}put} 
  \Input{a weight estimation $\vec w_1\in\W$, a step size $\alpha>0$, a sequence $x^n$ over $\X$,\\
    ~and a sequence $\mat P^n$ of probability matrices over $\P_+$
  }
  \Output{a codeword for $x^n$ of length $\ell(x^n,\mixogd(\vec w_1,\alpha,x^n, \mat P^n))$}
  \BlankLine
  \For{$k\gets1$ \KwTo $n$}{
    compute $p \gets \mix(\vec w_k, \mat P_k)$ and emit a codeword for $x_k$ sized $\ell(x_k, p)$ bits\;
    $\vec w_{k+1}\gets \proj(\vec w_k - \alpha\nabla_{\vec w} \ell(x_k, \mix(\vec w,\mat P_k)) \rvert_{\vec w=\vec w_k}; \W)$\;\label{line:gd_update}
  }
  \caption{$\mixogd(\vec w_1,\alpha,x^n,\mat P^n)$}\label{alg:mixgd}  
\end{algorithm}

\begin{definition}
A mixture $\mix : (\vec w, \mat P) \mapsto p$ maps a probability matrix $\mat P$ over $\P_+$, given a parameter vector $\vec w$ drawn from the parameter space $\W$, to a mixture distribution $p\in\P_+$. The shorthand $\mix(x; \vec w, \mat P)$ is for $p(x)$ where $p = \mix(\vec w, \mat P)$.
\end{definition}

\begin{definition}\label{def:lin}
For weight (parameter) vector $\vec w\in\S$ and probability matrix $\mat P$ over $\P_+$ the linear mixture $\lin$ is defined by $\lin(x; \vec w, \mat P) := \vec w\T \vec p(x)$.
\end{definition}

\begin{definition}
For weight (parameter) vector $\vec w\in\RR^m$ and probability matrix $\mat P$ over $\P_+$ the geometric mixture $\geo$ is defined by $\geo(x; \vec w, \mat P) := \prod_{i=1}^m p_i(x)^{w_i} / \sum_{y\in\X} \prod_{i=1}^m p_i(y)^{w_i}.$
\end{definition}

\begin{observation}\label{obs:geo}
If $\vec l(x) := -\log \vec p(x)$, then $\geo(x; \vec w, \mat P) = \powlw{x} / \sum_{y\in\X} \powlw{y}.$
\end{observation}

\noindent
In the following we will draw heavily on the alternate expression for $\geo(x; \vec w, \mat P)$ given in Observation \ref{obs:geo}. This expression simplifies some of the upcoming calculations. Furthermore, let
\begin{align}
  & \ell(x^n, \mixogd(\vec w_1, \alpha, x^n, \mat P^n)) := \sum_{k=1}^n \ell(x_k, \mix(\vec w_k, \mat P_k)) \text{ (for $\vec w_k$ see Algorithm \ref{alg:mixgd})},\\
  & \ell(x^n\hspace{-2pt}, \mat P^n\hspace{-2pt}, \vec w, \mix) \hspace{-2pt}:=\hspace{-4pt} \sum_{k=1}^n \ell(x_k, \mix(\vec w, \mat P_k)) \text{ and }
  \ell^*(x^n\hspace{-2pt}, \mat P^n\hspace{-2pt}, \mix) \hspace{-2pt}:=\hspace{-2pt} \min_{\mathclap{\vec w\in\W}} \ell(x^n\hspace{-2pt}, \mat P^n\hspace{-2pt}, \vec w, \mix).
\end{align}

\section{Nice Mixtures and Code Length Bounds}\label{sec:nice_mixtures}

\paragraph{Nice mixtures.}
We now introduce a class of especially interesting mixtures. We call such mixtures \emph{nice}. A nice mixture satisfies a couple of properties that allow us to derive bounds on the code length of combining such a mixture with \ac{OGD} for parameter estimation (e.g. weight estimation). These properties have been chosen carefully, s.t. linear and geometric mixtures fall into the class of nice mixtures (see Section \ref{sec:geo_lin_bounds}).

\begin{definition}\label{def:nice}
A mixture $\mix$ is called \emph{nice} if
\begin{enumerate}
  \item the parameter space $\W$ is a non-empty, compact and convex subset of $\RR^m$,\label{it:niceA}
  \item $\ell(x, \mix(\vec w, \mat P))$ is convex in $\vec w\in\W$ for all $\mat P$ over $\P^+$ and all $x\in\X$,\label{it:niceB}
  \item $\ell(x, \mix(\vec w, \mat P))$ is differentiable by $\vec w$ for all $\mat P$ over $\P^+$ and all $x\in\X$ and\label{it:niceC}
  \item there exists a constant $a>0$ s.t. $\lvert \nabla_{\vec w}\ell(x, \mix(\vec w, \mat P)) \rvert^2 \leq a\cdot \ell(x, \mix(\vec w, \mat P))$ for all $\vec w\in\W$, $\mat P$ over $\P^+$ and $x\in\X$.\label{it:niceD}
\end{enumerate}
\end{definition}

\begin{remark}
Properties 1 to 3 are similar to the assumptions made in \cite{zinkevich}, Property 4 differs. This allows us to obtain meaningful bounds on $\ell(x^n, \mixogd(\vec w_1, \alpha, x^n, \mat P^n))$ when $\alpha$ is independent of $n$, as \cite{bianchi99, nnloss} show.
\end{remark}

\paragraph{Bounds on the Code Length for \ac{OGD}.} Algorithm \ref{alg:mixgd} illustrates an online algorithm for mixture-based statistical data compression which employs a mixture $\mix$. We want to analyze the algorithm in terms of the number of bits required to encode a sequence when $\mix$ is nice. We strive to show that in some sense the code length produced by Algorithm \ref{alg:mixgd} is not much worse than a desirable competing scheme. At first we choose the code length produced by the best static weight vector $\vec w^* =\arg \min_{\vec w\in\W} \ell(x^n, \mat P^n, \vec w, \mix)$ as the competing scheme.

\begin{proposition}\label{prop:gd_bound}
Algorithm \ref{alg:mixgd} run with a nice mixture $\mix$, initial weight vector $\vec w_1\in\W$ and step size $\alpha=2(1-b^{-1})/a$ for $b>1$ (the constant $a$ is due to Definition \ref{def:nice}, Property \ref{it:niceD}) satisfies
\begin{align}
  \ell(x^n, \mixogd(\vec w_1, \alpha, x^n, \mat P^n)) ~\leq~ b \cdot \ell^*(x^n, \mat P^n, \mix) + \frac{a}{4} \frac{b^2}{b-1}\cdot\lvert \vec w_1-\vec w^*\rvert^2, \label{eq:gd_bound}
\end{align}
where $\vec w^*$ minimizes $\ell(x^n, \mat P^n, \vec w, \mix)$, for all $x^n$ over $\X$ and all $\mat P^n$ over $\P_+$.
\end{proposition}
\begin{prf}
For brevity we set $\ell_k(\vec w) := \ell(x_k, \mix(\vec w, \mat P_k))$. As in \cite{nnloss}, for arbitrary $\vec w\in\W$, we first establish a lower bound on
\begin{align}
  \lvert \vec w_k-\vec w\rvert^2-\lvert \vec w_{k+1}-\vec w \rvert^2
  &= \lvert \vec w_k-\vec w\rvert^2-\lvert \proj(\vec w_k-\alpha \nabla \ell_k(\vec w_k);\W)-\vec w \rvert^2.
\intertext{For $\vec v\in\RR^m$ and $\vec w\in\W$ it is well-known \cite{zinkevich}, that $\lvert \proj (\vec v; \W) - \vec w \rvert\leq \lvert\vec v-\vec w\rvert$, i.e.}
  \lvert \vec w_k-\vec w\rvert^2-\lvert \vec w_{k+1}-\vec w \rvert^2
  &\geq \lvert \vec w_k-\vec w\rvert^2 -\lvert (\vec w_k-\vec w) - \alpha \nabla \ell_k(\vec w_k) \rvert^2 \\
  &= 2\alpha\nabla \ell_k(\vec w_k)\T(\vec w_k-\vec w)-\alpha^2\lvert\nabla \ell_k(\vec w_k)\rvert^2.
\intertext{Since $\mix$ is nice, $\ell_k(\vec w)$ is convex (due to Definition \ref{def:nice}, Property \ref{it:niceB}) and we have $\ell_k(\vec v)- \ell_k(\vec w)\leq \nabla \ell_k(\vec v)\T(\vec v-\vec w)$ for any $\vec w, \vec v\in\W$. We deduce}  
  \lvert \vec w_k-\vec w\rvert^2-\lvert \vec w_{k+1}-\vec w \rvert^2
  &\geq 2\alpha(\ell_k(\vec w_k)-\ell_k(\vec w))-\alpha^2\lvert\nabla \ell_k(\vec w_k)\rvert^2 \\
  &\geq 2\alpha(\ell_k(\vec w_k)-\ell_k(\vec w))-a \alpha^2\ell_k(\vec w_k),\label{eq:progress_inv}
\end{align}
the last inequality follows from Definition \ref{def:nice}, Property \ref{it:niceD}. Next, we sum the previous inequality over $k$ to obtain (the sum telescopes)
\begin{align}
  \alpha(2-a\alpha)\sum_{k=1}^n\left(\ell_k(\vec w_k)\right) - 2\alpha\sum_{k=1}^n \ell_k(\vec w)
	&\leq \sum_{k=1}^n \lvert \vec w_k-\vec w\rvert^2-\lvert \vec w_{k+1}-\vec w \rvert^2 
	 \leq \lvert\vec w_1-\vec w\rvert^2,
\intertext{which we solve for the first sum:}
  \sum_{k=1}^n \ell_k(\vec w_k)
    &\leq \frac{2}{2-a\alpha} \sum_{k=1}^n\left(\ell_k(\vec w)\right) + \frac{\lvert\vec w_1-\vec w\rvert^2}{\alpha(2-a\alpha)}. \label{eq:gd_bound1}
\end{align}
Since this holds for any $\vec w$, it must hold for $\vec w=\vec w^*$, too. By the definition of $\ell_k(\vec w)$ we have 
$\sum_{k=1}^n \ell_k(\vec w_k) = \ell(x^n, \mixogd(\vec w_1, \alpha, x^n, \mat P^n))$ and  $\sum_{k=1}^n \ell_k(\vec w) = \ell(x^n,\mat P^n,\vec w, \mix)$.
Our choice of $\alpha$ gives \eqref{eq:gd_bound}. 
\end{prf}

\begin{remark}
The technique of using a progress invariant (c.f. \eqref{eq:progress_inv}) in the previous proof is adopted from the machine learning community, see \cite{bianchi99,nnloss}.
These two papers assume that the domain of the parameter (weight) vector $\vec w$ is unbounded. Techniques of \cite{zinkevich} allow us to overcome this limitation.
Proposition \ref{prop:gd_bound} generalizes the analysis of online regression of \cite{bianchi99} to prediction functions $f(\vec w, \vec z)$ ($\vec z$ is the input vector for a prediction) instead of $f(\vec w\T\vec z)$ when the domain of $\vec w$ is restricted.
\end{remark}

The previous proposition is good news. The number of bits required to code any sequence will be within a multiplicative constant $b$ of the code length generated by weighting with an optimal fixed weight vector, $\ell^*(x^n, \mat P^n, \mix)$, plus an $O(1)$ term. At the expense of increasing the $O(1)$ term we can set the multiplicative constant $b$ arbitrarily close to $1$. Note that the $O(1)$-term originates in the inaccuracy of the initial weight estimation $\lvert \vec w_1-\vec w^*\rvert$ (see \eqref{eq:gd_bound}) and as $b$ approaches $1$, the step size $\alpha$ approaches zero. Hence the $O(1)$ term in \eqref{eq:gd_bound} penalizes a slow movement away from $\vec w_1$. A high proximity of $\vec w_1$ to the optimal weight vector $\vec w^*$ damps this penalization. We now make two key observations, which allow us to greatly strengthen the result of Proposition \ref{prop:gd_bound}.

\begin{observation}\label{obs:b}
From the previous discussion we know that the significance of the $O(1)$ term vanishes as $\ell^*(x^n, \mat P^n, \mix)$ grows. We can allow small values of $b$ for large values of $n$, i.e., $b$ may depend on $n$. Thus we choose $b=1+f(n)$ where, $f(n)$ decreases, and obtain 
\begin{multline}
  \ell(x^n, \mixogd(\vec w_1, \alpha, x^n, \mat P^n)) \\ \leq \ell^*(x^n, \mat P^n, \mix) + \ell^*(x^n, \mat P^n, \mix)\cdot f(n) + \frac{a (1+f(1))^2 \lvert \vec w_1-\vec w^*\rvert^2}{4} \cdot \frac{1}{f(n)}.
\end{multline}
If $\ell^*(x^n, \mat P^n, \mix)$ is $O(n)$ (i.e., $\mix(x; \vec w, \mat P)$ is bounded below by a constant, which is a natural assumption) then the rightmost two terms on the previous line are $O(n\cdot f(n) + f(n)^{-1})$ (since by Definition \ref{def:nice}, Property \ref{it:niceA}, $\lvert\vec w_1-\vec w^*\rvert$ is $O(1)$) and represent the number of bits wasted by $\mixogd$ w.r.t. $\ell^*(x^n, \mat P^n, \mix)$. Clearly the rate of growth is minimized in the $O$-sense if we choose $f(n)=n^{-1/2}$, i.e. $\ell(x^n, \mixogd(\vec w_1, \alpha, x^n, \mat P^n))\leq \ell^*(x^n, \mat P^n, \mix) + O(n^{1/2})$. The average code length excess of $\mixogd$ over $\ell^*(x^n, \mat P^n, \mix)$ vanishes asymptotically.
\end{observation}

\begin{observation}\label{obs:split}
The state of $\mixogd$ right after step $k$ is captured completely by the single weight vector $\vec w_{k+1}$. Hence we can view running $\mixogd(\vec w_1, \alpha, x^n, \mat P^n)$ as first executing $\mixogd(\vec w_1, \alpha, x^k, \mat P^k)$ and running $\mixogd(\vec w_{k+1}, \alpha, x_{k+1}^n, \mat P_{k+1}^n)$ afterwards. The code lengths for these procedures match for all $1\leq k < n$:
\begin{multline}
  \ell(x^n, \mixogd(\vec w_1, \alpha, x^n, \mat P^n)) \\ = \ell(x^k, \mixogd(\vec w_1, \alpha, x^k, \mat P^k)) + \ell(x^n_{k+1}, \mixogd(\vec w_{k+1}, \alpha, x_{k+1}^n \mat P^n_{k+1})).
\end{multline}
\end{observation}

\noindent
Given the previous observations as tools of trade we now enhance Proposition \ref{prop:gd_bound}.

\begin{theorem}\label{thm:strong_gd_bound}
We consider sequences $t_1=1<t_2<\dots<t_s<t_{s+1}=n+1$ of integers for $1\leq s\leq n$.
Let $\ell^*(i,j,\mix) := \ell^*(x_i^j, \mat P_i^j, \mix)$.
For all $x^n\in\X^n$, all $\mat P^n$ over $\P_+$, any nice mixture $\mix$ and any $\vec w_1\in\W$ Algorithm \ref{alg:mixgd} satisfies:
\begin{enumerate}
  \item\label{it:strong_gd_boundA}
    If $\alpha = 2(1-b^{-1})/a$, where $b>1$, then
    \begin{align}
      \hspace{-12pt}
      \ell(x^n, \mixogd(\vec w_1, \alpha, x^n, \mat P^n)) 
       \leq \min_{s, \,t_2,\dots,t_s} \left[ \frac{a b^2\lvert \W\rvert^2}{4(b-1)}  s + b \sum_{i=1}^s \ell^*(t_i, t_{i+1}{-}1, \mix) \right]. \label{eq:strong_gd_boundA}
    \end{align}
  \item\label{it:strong_gd_boundB}
    If $\alpha = 2/a \cdot (1+n^{1/2})^{-1}$ (i.e., $b=1+n^{-1/2}$) and $\ell^*(x^n, \mat P^n, \mix)\leq c\cdot n$ holds for a constant $c>0$, all $x^n$ over $\X$ and all $\mat P^n$ over $\P_+$ then
    \begin{align}
     \hspace{-6pt}
     \ell(x^n, \mixogd(\vec w_1, \alpha, x^n, \mat P^n))
      \leq \hspace{-2pt}\min_{s, \,t_2,\dots,t_s} \hspace{-3pt}\left[ \left(a s \lvert W\rvert^{\mathclap{~2}} + c\right) \sqrt n + \hspace{-2pt}\sum_{i=1}^s \ell^*(t_i, t_{i+1}{-} 1, \mix) \right]. \label{eq:strong_gd_boundB}
   \end{align}  
\end{enumerate}
\end{theorem}
\begin{prf}
We start proving \eqref{eq:strong_gd_boundA}. First, we define $\ell_k(\vec w) := \ell(x_k, \mix(\vec w, \mat P_k))$. By Observation \ref{obs:split} for any $1\leq s\leq n$ and $t_1=1<t_2<\dots<t_s<t_{s+1}=n+1$ we may write
\begin{align}
  \ell(x^n, \mixogd(\vec w_1, \alpha, x^n, \mat P^n))   
  &= \sum_{i=1}^s \ell(x_{{t_i}}^{{t_{i+1}}-1}, \mixogd(\vec w_{{t_i}}, \alpha, x_{{t_i}}^{{t_{i+1}}-1}, \mat P_{{t_i}}^{{t_{i+1}}-1})) \\
  &\leq \frac{a \lvert \W \rvert^2}{4} \frac{b^2}{b-1}\cdot s + b \sum_{i=1}^s \ell^*(t_i, t_{i+1}-1,\mix). \label{eq:strong_gd_bound1}
\end{align}
For the last step we used Proposition \ref{prop:gd_bound}, the definition of $\ell^*(t_i, t_{i+1}-1,\mix)$ and Definition \ref{def:nice}, Property \ref{it:niceA} which implies that $\lvert \vec v-\vec w\rvert\leq \lvert \W\rvert$ for any $\vec v, \vec w\in\W$. Since this holds for arbitrary $s$ and $t_2, \dots, t_s$ we can take the minimum over the corresponding entities, which gives \eqref{eq:strong_gd_boundA}.

Now we turn to \eqref{eq:strong_gd_boundB}. The choice $b = 1+n^{-1/2}$ follows from Observation \ref{obs:b}. We combine $b^2/(b-1)\leq 4 n^{1/2}$ (by the choice of $b$) with $\ell^*(x^n, \mat P^n, \mix)\leq c \cdot n$, i.e. $\ell^*(i,j,x^n)\leq c\cdot (j-i+1)$ for $j\geq i$ in the r.h.s. of \eqref{eq:strong_gd_bound1} to yield
\begin{align}
  \ell(x^n, \mixogd(\vec w_1, \alpha, x^n, \mat P^n))
  &\leq a \lvert \W \rvert^2 s \cdot n^{1/2} + (1+n^{-1/2}) \sum_{i=1}^s \ell^*(t_i, t_{i+1}-1,\mix)  \\
  &\leq \left(a \lvert \W \rvert^2 s + c\right) \cdot n^{1/2} + \sum_{i=1}^s \ell^*(t_i, t_{i+1}-1,\mix).
\end{align}
As in the proof of \eqref{eq:strong_gd_boundA} we take the minimum over $s$ and $t_2, \dots, t_s$, which gives \eqref{eq:strong_gd_boundB}.
\end{prf}

The previous theorem gives much stronger bounds than Proposition \ref{prop:gd_bound}, since the competing scheme is a sequence of weight vectors with a total code length of $\ell^*(t_1, t_2-1, \mix) + \dots + \ell^*(t_s, t_{s+1}-1, \mix)$, where the $i$-th weight vector minimizes the code length of the $i$-th subsequence $x_{t_i}\dots x_{t_{i+1}-1}$ of $x^n$. By \eqref{eq:strong_gd_boundA} the performance of Algorithm \ref{alg:mixgd} is  within a multiplicative constant $b>1$ of the performance of any competing scheme (since in \eqref{eq:strong_gd_boundA} we take the minimum over all competing schemes) plus an $O(s)$-term, when $\alpha$ is independent of $n$. The $O(s)$ term penalizes the complexity of a competing predictor (the number $s$ of subsequences). When $\alpha$ depends on $n$ (c.f. \eqref{eq:strong_gd_boundB}) we can reduce the multiplicative constant to $1$ at the expense of increasing the penalty term to $O(s \sqrt n)$, i.e. Algorithm \ref{alg:mixgd} will asymptotically perform not much worse than any such competing scheme with $s=o(\sqrt n)$ subsequences.


\section{Bounds for Geometric and Linear Mixtures}\label{sec:geo_lin_bounds}

\paragraph{Geometric and Linear Mixtures are Nice.}

We can only apply the machinery of the previous section to geometric and linear mixtures if they fall into the class of nice mixtures. Since the necessary conditions have been chosen carefully, this is the case:
  
\begin{lemma}\label{lem:geo_nice}
The geometric mixture $\geo(\vec w, \mat P)$ is nice for $\vec w\in\W$, if $\W$ is a compact and convex subset of $\RR^m$. Property \ref{it:niceD} of Definition \ref{def:nice} is satisfied for $a\geq \frac{m}{\log(e)} \log^2\left(\pmax/\pmin\right)$.
\end{lemma}

\begin{lemma}\label{lem:lin_nice}
The linear mixture $\lin(\vec w, \mat P)$ is nice. Property \ref{it:niceD} of Definition \ref{def:nice} is satisfied for $a\geq m \log^2(e)  \frac{\pmax^2}{\pmin^2 \log(1/\pmin)}$.
\end{lemma}

\noindent
Before we prove these two lemmas we give two technical results. The proofs of the lemmas below use standard calculus, we omit them for reasons of space.

\begin{restatable}{lemma}{fzgeo}\label{lem:fz_geo}
For $0<z<1$ the function $f(z):=-\frac{\ln z}{1-z}$ satisfies $f(z)\geq1$.
\end{restatable}

\begin{restatable}{lemma}{fzlin}\label{lem:fz_lin}
For $0<a\leq z\leq 1-a$ the function $f(z) := -z^2\ln z$ satisfies $f(z)\geq f(a)$.
\end{restatable}

\noindent
Now we are ready to prove Lemma \ref{lem:geo_nice} and Lemma \ref{lem:lin_nice}. 

\begin{prf}[Proof of Lemma \ref{lem:geo_nice}]
Let $p(x; \vec w) := \geo(x; \vec w, \mat P)$ and $\ell(\vec w) := \ell(x, \geo(\vec w, \mat P))$.
To show the claim we must make sure that properties \ref{it:niceA}-\ref{it:niceD} of Definition \ref{def:nice} are met. By the constraint on $\W$ Property \ref{it:niceA} is satisfied. Property \ref{it:niceB} was shown in \cite[Section 3.2]{cm_dcc2012}. To see that Property \ref{it:niceC} holds, we set $c:=\sum_{y\in\X} \powlw y$ and compute
\begin{align}
  \nabla \ell(\vec w) 
     & = \nabla_{\vec w} \left(\vec w\T \vec l(x)+\log c\right) 
     = \vec l(x) - \sum_{\mathclap{y\in\X}} \frac{\powlw y}{c} \cdot\vec l(y), \\
\intertext{which is (by the definition of $\geo$)}
  \nabla \ell(\vec w) 
    &= \nabla_{\vec w} \ell(x, \geo(\vec w, \vec P))
    = \sum_{y\neq x}\geo(y; \vec w, \mat P) \cdot \left(\vec l(x) - \vec l(y)\right). \label{eq:geo_grad}
\end{align}
Clearly \eqref{eq:geo_grad} is well-defined for the given range of $\vec w$ and $\vec P$. For Property \ref{it:niceD} we bound $\lvert \nabla \ell(\vec w) \rvert^2 / \ell(\vec w)$ from above by a constant; $a$ takes at least the value of this constant. We obtain 
\begin{align}
  \lvert\nabla \ell(\vec w)\rvert^2     
    &\leq \sum_{y\neq x} p(y; \vec w) \lvert\vec l(x)-\vec l(y)\rvert^2
    = \sum_{y\neq x} p(y; \vec w) \sum_{i=1}^m \log^2\frac{p_i(y)}{p_i(x)} \\
    &\leq \sum_{y\neq x}p(y; \vec w) m \log^2\frac{\pmax}{\pmin}
    = (1-p(x; \vec w)) m \log^2\frac{\pmax}{\pmin} \text{ and }\\
  \frac{\lvert\nabla\ell(\vec w)\rvert^2}{\ell(\vec w)}
    &\leq \frac{ (1-p(x; \vec w)) m \log^2\left(\frac{\pmax}{\pmin}\right)}{-\log p(x; \vec w)} 
    \leq \frac{m\log^2\left(\frac{\pmax}{\pmin}\right)}{\log(e)} \cdot \left[ \inf_{0<z<1} -\frac{\ln z}{1-z}\right]^{-1}.
\end{align}
By Lemma \ref{lem:fz_geo} the infimum is at least $1$. This yields the claimed lower bound on $a$.
\end{prf}

\begin{remark}
It is interesting to note that we can express $\nabla_{\vec w} \ell(x, \geo(\vec w, \mat P))$ (see \eqref{eq:geo_grad}) in terms of information theoretic quantities (for the basic notation see, e.g. \cite{eoit}). The $i$-th component is
\begin{align}
  &-\log p_i(x) - \sum_{y\in\X} \geo(y; \vec w, \mat P) (-\log p_i(y)) \\[-6pt]
  &\quad = -\log p_i(x) - \sum_{y\in\X}\geo(y; \vec w, \mat P) \left[\log\left(\frac{1}{\geo(y; \vec w, \mat P)}\right) + \log \frac{\geo(y; \vec w, \mat P)}{p_i(y)}\right] \\
  &\quad = -\log p_i(x) - (H(\geo(\vec w, \mat P)) + D(\geo(\vec w, \mat P)\parallel p_i)).
\end{align}
If we now ignore possible constraints on the weight vector $\vec w$ then for some character $x$ a minimizer of $\min_{\vec w} \ell(x, \geo(\vec w, \mat P))$ satisfies $H(\geo(\vec w, \mat P)) + D(\geo(\vec w, \mat P)\parallel p_i) = -\log p_i(x)$ for all $1\leq i\leq m$. In effect the weight vector $\vec w$ is chosen s.t. there is an equilibrium: The code length $-\log p_i(x)$ matches the average code length of coding a symbol drawn from the source distribution $\geo(\vec w, \mat P)$ with the model distribution $p_i$.
\end{remark}

\begin{prf}[Proof of Lemma \ref{lem:lin_nice}]
Again we set $p(x; \vec w) := \lin(x; \vec w, \mat P)$ and $\ell(\vec w) := \ell(x, \lin(\vec w, \mat P))$ and proceed analogously to the proof of Lemma \ref{lem:geo_nice}.
By Definition \ref{def:lin} we have $\vec w\in\S$, Property \ref{it:niceA} is met, and in \cite[Section 4.2]{cm_dcc2012} we showed that Property \ref{it:niceB} is met, as well. The gradient
\begin{align}
  &\quad\nabla \ell(\vec w)
  = \nabla_{\vec w} \ell(x, \lin(\vec w, \mat P))
  = -\nabla_{\vec w} \log \vec w\T \vec p(x)
  = -\log(e)\frac{\vec p(x)}{p(x; \vec w)} \label{eq:lin_grad}\\
\intertext{is well-defined for the given range of $\vec w$ and $\mat P$, so Property \ref{it:niceC} is fulfilled. We observe that}
&\quad\frac{\lvert\nabla\ell(\vec w)\rvert^2}{\ell(\vec w)}
  \leq \frac{m\log^2(e)\pmax^2}{p(x; \vec w)^2 (-\log p(x; \vec w))}
  \leq m\log(e)\pmax^2\cdot\left[ \inf_{c\leq z\leq d} -z^2 \ln z \right]^{-1} \label{eq:lin_nice1} \\
\intertext{where $c=\pmin\leq p(x; \vec w)\leq \pmax\leq d=1-\pmin$. We used $\pmax\leq 1-\pmin$, since}
  &\pmax = \max_{1\leq i\leq m}\max_{x\in\X} p_i(x) \leq \max_{1\leq i\leq m} \left(1-\min_{x\in\X} p_i(x)\right) = 1-\min_{1\leq i\leq m}\min_{x\in\X} p_i(x) = 1-\pmin,
\end{align}
to apply Lemma \ref{lem:fz_lin} to bound the rightmost factor in \eqref{eq:lin_nice1} from above by $[-\pmin^2\ln\pmin]^{-1}$. The resulting constant on the r.h.s. of \eqref{eq:lin_nice1} is a lower bound on $a$. The proof is done.
\end{prf}

\paragraph{Upper bounds on the Code Length.}
At this point we can combine Theorem \ref{thm:strong_gd_bound} with Lemmas \ref{lem:geo_nice} and \ref{lem:lin_nice} to obtain code length bounds on Algorithm \ref{alg:mixgd} for $\lin$ and $\geo$. The discussion in Section \ref{sec:nice_mixtures} on nice mixtures coupled with Algorithm \ref{alg:mixgd} applies to $\lin$ and $\geo$ as well.

\begin{theorem}\label{thm:bounds}
Let $x^n\in\X^n$, let $\mat P^n$ be a sequence of probability matrices over $\P_\eps$ where $\eps=2^{-B}$ for $1\leq B<\infty$ and let $\ell^*(k,l,\best) := \min_{1\leq i\leq m} \ell(x_k^l, \lin(\vec e_i,\mat P_k^l))$ be the code length of the best single model for $x_k^l$. We consider sequences $t_1=1<t_2<\dots<t_s<t_{s+1}=n+1$ of integers where $1\leq s\leq n$. For $\mix=\lin$ and $\mix=\geo$ where $\W=\S$ Algorithm \ref{alg:mixgd} satisfies the bounds in Table \ref{tab:bounds} for the given step sizes for all $\vec w_1\in\S$.
\end{theorem}
\begin{prf}
For the sake of simplicity we set $\linogd(\alpha) := \linogd(\vec w_1, \alpha, x^n, \mat P^n)$. We start by proving row $1$ in Table \ref{tab:bounds}. By Lemma \ref{lem:lin_nice} we can use Theorem \ref{thm:strong_gd_bound}, Equation \eqref{eq:strong_gd_boundA} with $\mix=\lin$, $b=2$ and $\W=\S$ where $\lvert\S\rvert^2\leq 2$ which gives
\begin{align}
  &\alpha = \frac{1}{a} ~\text{ and }~ \ell(x^n, \linogd(\alpha)) \leq  2 a s + 2 \sum_{i=1}^s \ell^*(t_i, t_{i+1}-1, \lin) \label{eq:bounds1}
\intertext{for any $s, t_2, \dots, t_s$. Observe that}
  & \ell^*(k, l, \lin) = \min_{\vec w\in\S} \ell(x_k^l, \lin(\vec w, \mat P_k^l)) \leq \min_{1\leq i\leq m} \ell(x_k^l, \lin(\vec e_i, \mat P_k^l)) = \ell^*(k,l,\best)\label{eq:bounds2}\\
\intertext{and by Lemma \ref{lem:lin_nice} we can choose}
  & a = \frac{17m 4^B}{8 B} \cdot f(n) \geq \frac{17 m}{8} \frac{1}{\eps^2\log(1/\eps)} \geq m \log^2(e)  \frac{\pmax^2}{\pmin^2 \log(1/\pmin)}. \label{eq:bounds3} 
\intertext{for some $f(n)\geq1$. We set $f(n)=1$ and combine \eqref{eq:bounds2} and \eqref{eq:bounds3} with \eqref{eq:bounds1} to yield}
  & \alpha = \frac{8 B}{17m 4^B} ~\text{ and }~ \ell(x^n, \linogd(\alpha)) \leq  \frac{17m s 4^B}{4 B} + 2 \sum_{i=1}^s \ell^*(t_i, t_{i+1}-1, \best).
\intertext{Finally we can take the minimum over $s, t_2, \dots, t_s$, since these were arbitrary, which gives the claim. Now we advance to Table \ref{tab:bounds}, row $2$. Again, by Lemma \ref{lem:lin_nice} we use Theorem \ref{thm:strong_gd_bound}, Equation \eqref{eq:strong_gd_boundB} with $\mix=\lin$, $c=-\log \eps=B$ and $\W=\S$ which gives}
  &\alpha = \frac{2/a}{1+\sqrt n} \,\text{ and }\, \ell(x^n, \linogd(\alpha)) \leq  (2 a s+B)\sqrt n + \sum_{i=1}^s \ell^*(t_i, t_{i+1}-1, \lin) \label{eq:bounds4} \\
\intertext{for any $s, t_2, \dots, t_s$. We now choose $a$ as in \eqref{eq:bounds3} with $1\leq f(n) = \frac{2\sqrt n}{1+\sqrt n}\leq 2$, to get}
 &(2 a s + B) = \frac{17 m s 4^B}{4B} f(n)+B \leq (17 m s +1)\frac{4^B}{2B}\leq \frac{35 m s 4^B}{4B} \label{eq:bounds41} \\
\intertext{for the constant on the r.h.s. of \eqref{eq:bounds4}.We combine \eqref{eq:bounds2} and \eqref{eq:bounds41} with \eqref{eq:bounds4} to yield}
  &\alpha = \frac{8 B/\sqrt n}{17m 4^B} ~\text{ and }~ \ell(x^n, \linogd(\alpha)) \leq \frac{35 m s 4^B}{4B}\sqrt n + \sum_{i=1}^s \ell^*(t_i, t_{i+1}-1, \best).
\intertext{Again, taking the minimum over $s, t_2, \dots, t_s$ finishes the proof. The bounds of Table \ref{tab:bounds} rows 3 and 4 follow analogously by the choice of $f(n)=1$ (row 3) and $f(n)=\frac{2\sqrt n}{1 +\sqrt n}$ (row 4) and}
  &a = \frac{7 m B^2}{10}\cdot f(n) \geq \frac{7m}{10}\log^2\frac1\eps \geq \frac{m\log^2(\pmax/\pmin)}{\log e} \text{ and by} \\
  &\ell^*(k, l, \geo) = \min_{\vec w\in\S} \ell(x_k^l, \geo(\vec w, \mat P_k^l)) \leq \min_{1\leq i\leq m} \ell(x_k^l, \geo(\vec e_i, \mat P_k^l)) = l^*(k,l,\best)\\
\intertext{and using $\ell^*(x^n,\geo(\mat P^n)) \leq B\cdot n$ (a premise of Theorem \ref{thm:strong_gd_bound}, Item \ref{it:strong_gd_boundB}), since for all $\vec w\in\S$}
  &\geo(x; \vec w, \mat P) = \frac{\prod_{i=1}^m p_i(x)^{w_i}}{\sum_{y\in\X} \prod_{i=1}^m p_i(y)^{w_i}} \geq \prod_{i=1}^m p_i(x)^{w_i} \geq \pmin(x; \mat P) = \eps = 2^{-B} \label{eq:bounds5}
\end{align}
and consequently $\ell(x, \geo(\vec w, \mat P))\leq B$.
\end{prf}
\begin{table}[h]
  \centering
  \caption{Code length bounds of Algorithm \ref{alg:mixgd} for $\mix=\lin$ and $\mix=\geo$ where $\W=\S$.}
  \vspace*{6pt}
  \begin{tabular}{rlcc}
  \toprule
  & $\mix$ & $\alpha$ & $\ell(x^n, \mixogd(\vec w_1, \alpha, x^n, \mat P^n)) \leq \min_{\textstyle s, \,t_2,\dots,t_s}$ of \dots \\ \midrule
  1 & $\lin$ &
    $\displaystyle \frac{8 B}{17 m 4^{B}}$ &
    \makebox[0.45\textwidth][l]{$\displaystyle 2 \sum_{i=1}^s \ell^*(t_i, t_{i+1}-1,\best) + \frac{17 m s 4^{B}}{4 B}$} \\[12pt]
  2 & &
    $\displaystyle \frac{8 B/\sqrt{n}}{17 m 4^{B}}$ &
    \makebox[0.45\textwidth][l]{$\displaystyle  \phantom{2} \sum_{i=1}^s \ell^*(t_i, t_{i+1}-1,\best) + \frac{35 m s 4^{B}}{4B}\sqrt{n}$} \\[12pt] 
  3 & $\geo$ &
    $\displaystyle \frac{10}{7 m B^2}$ &
    \makebox[0.45\textwidth][l]{$\displaystyle  2 \sum_{i=1}^s \ell^*(t_i, t_{i+1}-1,\best) + \frac{7 m s B^2}{5}$} \\[12pt]
  4 & &
    $\displaystyle \frac{10/\sqrt n}{7 m B^2}$ & 
    \makebox[0.45\textwidth][l]{$\displaystyle \phantom{2}\sum_{i=1}^s \ell^*(t_i, t_{i+1}-1,\best) + \frac{19 m s B^2}{10} \sqrt n$} \\
  \bottomrule
  \end{tabular}
  \label{tab:bounds}
\end{table}
\begin{remark}
In the previous proof \eqref{eq:bounds5} shows that $\geo(x; \vec w, \mat P)\geq \pmin(x; \mat P)$ when $\vec w\in\S$, just as $\lin$. Subsequently $\geo$ cannot use more bits than $\lin$ to encode a single symbol in the worst case. In the best case $\geo(x; \vec w, \mat P)$ uses \emph{at most} as much bits as $\lin$, since
\begin{align}
  \max_{\vec w\in\S} \geo(x; \vec w, \mat P)) \geq \max_{1\leq i\leq m} \geo(x; \vec e_i, \mat P) = \pmax(x; \mat P) = \max_{\vec w\in\S} \lin(x; \vec w, \mat P).
\end{align}
There exist situations where $\displaystyle \max_{\vec w\in\S}\geo(x; \vec w, \mat P)>\max_{\vec w\in\S}\lin(x; \vec w, \mat P)$, see Example \ref{ex:geo}.
\end{remark}

\begin{example}\label{ex:geo}
For an alphabet $\X = \{1, 2, \dots, N\}, N>2$, we consider $\geo(\vec w, \mat P)$ where $\vec w = (1/2~1/2)\T$ and $\mat P\T = (p_1(x)~p_2(x))_{x\in\X}$ s.t. for $0<\eps,q<1$ we have
\begin{align}
  &p_1(x) := \begin{cases} q &, x=1 \\ (1-q)\cdot(1-\eps) &, x=2 \\ \frac{(1-q)\cdot\eps}{N-2} &, \text{otherwise} \end{cases},~
  p_1(x) := \begin{cases} q &, x=1 \\ (1-q)\cdot(1-\eps) &, x=3 \\ \frac{(1-q)\cdot\eps}{N-2} &, \text{otherwise} \end{cases}, \\
\intertext{The mixture probability \geo(1; \vec w, \mat P) of the letter $1$ is}  
  &\qquad\frac{p_1(1)^{1/2} \cdot p_2(1)^{1/2} }{\sum_{y\in\X} p_1(y)^{1/2} \cdot p_2(y)^{1/2}}
  = q / \bigg[q + (1-q)  \underbrace{ \left(2\sqrt{\frac{\eps(1-\eps)}{N-2}} + \frac{N-3}{N-2}\eps\right) }_{\mathclap{=: f(\eps,N)}} \bigg],\\
\intertext{We now show, that for any $q$ there exists an $\eps$, such that $\geo(1; \vec w, \mat P)> \pmax(1; \mat P) = q$. Clearly, if $\geo(1; \vec w, \mat P)>q$ we must have $f(\eps,N)<1$. To observe this we bound $f(\eps,N)$ from above and give a possible choice for $\eps$.}
  &f(\eps,N)=2\sqrt{\frac{\eps(1-\eps)}{N-2}} + \frac{N-3}{N-2}\eps \leq 2\sqrt{\frac\eps{N-2}} + (N-3)\sqrt{\frac{\eps}{N-2}} = \frac{N-1}{\sqrt{N-2}}\cdot\sqrt{\eps}
\end{align}
If we choose $0<\eps<(N-2)/(N-1)^2$ it follows that $f(\eps,N)<1$ and $\geo(1; \vec w,\mat P) > q$.
\end{example}

Note that the bounds in Table \ref{tab:bounds}, rows 3 and 4 only translate to \ac{PAQ}7 if $\W=\S$. To obtain bounds for other weight spaces $\W$ we only need to substitute the approrpiate values for $\lvert\W\rvert$ and/or $c>0$ where $\ell(x, \geo(\vec w, \mat P))\leq c$ in the previous proof. E.g., if we have $-r\cdot\vec 1\leq \vec w\leq r\cdot \vec 1$ for $r>0$ then the penalization term of the bound in row 3 increases by a factor of $\lvert\W\rvert^2/\lvert\S\rvert^2 = m r^2$.

Veness \cite{veness12} gave a bound for linear mixtures using a non-\ac{OGD} weight estimation scheme which is identical to Table \ref{tab:bounds} row 2 except the penalty term, which is $O(s\log n)$ in place of $O(s\sqrt n)$. However our analysis is based on Theorem \ref{thm:strong_gd_bound} which applies to the strictly larger class of nice mixtures with a generic scheme for weight estimation. Clearly, more restrictions can pay off in tighter bounds, consequently we might obtain better bounds by taking advantage of the peculiarities of $\lin$ and $\geo$.

\section{Conclusion}\label{sec:conclusion}

In this work we obtained code length guarantees for a particular mixture-based adaptive statistical data compression algorithm. The algorithm of interest combines multiple model distributions via a mixture and employs \ac{OGD} to adjust the mixture parameters (typically model weights). As a cornerstone we introduced the class of nice mixtures and gave bounds on their code length in the aforementioned algorithm. Since, as we showed, linear and geometric mixtures are nice mixtures we were able to deduce code length guarantees for these two mixtures in the above data compression algorithm. Our results on geometric mixtures directly apply to \ac{PAQ}7, a special case of geometric mixtures, and provide the first analysis of \ac{PAQ}7.

We defer an exhaustive experimental study on linear and geometric mixtures to future research. A straightforward extension to Theorem \ref{thm:strong_gd_bound}, Item \ref{it:strong_gd_boundB} is to remove the dependence of the step size on the sequence length (which is typically not known in advance). This can be accomplished by using the ``doubling-trick'' \cite{plg} or a decreasing step size \cite{zinkevich}. Another interesting topic is whether geometric and/or linear mixtures have disjoint properties, which we can use to yield stronger bounds. This opposes our current approach, which we built on the (common) properties of a nice mixture.

\paragraph*{Acknowledgement.}
The author would like to thank Martin Dietzfelbinger, Michael Rink, Sascha Grau and the anonymous reviewers for valuable improvements to this work.

\bibliographystyle{plain}

\appendix
\section{Proof of Lemma \ref{lem:fz_geo} and Lemma \ref{lem:fz_lin}}
\fzgeo*
\begin{prf}
By the basic inequality $-\ln(z)\geq 1-z$ the claim follows.
\end{prf}

\fzlin*
\begin{prf}
First, we examine the derivative $f'(z) = -z(1+2\ln z)$ of $f$. Clearly, $f'(z)\geq0$ for $0< z<z_0:=1/\sqrt e$ and $f'(z)\leq0$ for $z_0\leq z\leq 1$. From $a\leq 1-a$ we conclude that $a\leq\frac12$. We have $f(z)\geq \min\{f(a), f(1-a)\}$ (by monotonicity) and it remains to show that $f(a)\leq f(1-a)$. Let $g(a) := f(a)/f(1-a)$ and observe that $g(a)$ increases monotonically for $0<a\leq\frac12$, i.e. $\frac{f(a)}{f(1-a)}=g(a)\leq g(\frac12) = 1$. Finally we argue that $g'(a)\geq0$ where
\begin{align}
g'(a) = \frac{a\ln(a)\ln(1-a)}{(a-1)^3 \ln^2(1-a)}\cdot\left[ -\frac{a}{\ln(1-a)}-\frac{1-a}{\ln a}-2\right].
\end{align}
Clearly, the left factor is negative for $0<a\leq\frac12$. The rightmost factor is at most 0, since by Lemma \ref{lem:fz_geo} we have $-\frac{a}{\ln(1-a)}\leq 1 / \inf_{0<z<1}\,-\frac{\ln z}{1-z} \leq 1$ (we substituted $z=1-a$) and $-\frac{1-a}{\ln a} \leq 1/\inf_{0<z<1}\,-\frac{\ln z}{1-z} \leq 1$, which concludes the proof.
\end{prf}


\end{document}